\documentstyle[12pt]{article}
\newcommand{\be}{\begin{equation}}
\newcommand{\ee}{\end{equation}}

\newcommand{\bea}{\begin{eqnarray}}
\newcommand{\eea}{\end{eqnarray}}
\newcommand{\ba}{\begin{array}}
\newcommand{\ea}{\end{array}}
\newcommand{\beqa}{\begin{eqnarray}}
\newcommand{\eeqa}{\end{eqnarray}}

\newcommand{\PL}[1]{Phys. Lett.\ {\bf #1}\ }

\newcommand{\PR}[1]{Phys. Rev.\ {\bf #1}\ }

\newcommand{\cL}{{\cal L}}

\newcommand{\half}{{1\over2}}

\newcommand{\D}{\delta}
\newcommand{\DE}{\Delta}
\newcommand{\G}{\gamma}

\newcommand{\eg}{{\it e.g. }}

\newcommand{\ssu}{$SU(2)_L\times SU(2)_R\times U(1)_{B-L}\,$}

\newcommand{\matr}{\left( \begin{array}}
\newcommand{\ematr}{\end{array} \right)}

\newcommand{\dis}{\displaystyle}

\begin{document}

\begin{titlepage}

\mbox{}\vspace*{-1cm}\hspace*{9cm}\makebox[7cm][r]{\large  HU-SEFT R 1993-17}
\vfill

\Large

\begin{center}
{\bf  Slepton pair production in $e^+e^-$ collision in supersymmetric
left-right
 model}

\bigskip
\normalsize
{${\rm K. Huitu,}^a\:\: {\rm J. Maalampi}^b\:\: {\rm and} \:\:  {\rm M.
Raidal}^{a,b}$\\[15pt] $^a${\it Research Institute for High Energy Physics,
University of Helsinki}\\$^b${\it Department of Theoretical Physics,
University of Helsinki}}

{December 1993}

\bigskip


\vfill

\normalsize

{\bf\normalsize \bf Abstract} \end{center}
\normalsize
 The pair production of  sleptons in
electron-positron collisions is investigated in a supersymmetric
left-right model. The cross section is found  considerably larger
than in the minimal supersymmetric version of the Standard Model
(MSSM) because of more  contributing graphs. A novel process is  a
doubly charged higgsino exchange in u-channel, which makes the
angular distribution of the final state particles and the final
state asymmetries to differ from those of the MSSM. It also allows
for the flavour non-diagonal final states $\tilde e\tilde\mu$,
$\tilde e\tilde\tau$ and $\tilde \mu\tilde\tau$, forbidden in the
MSSM. These processes also give  indirect information about
neutrino mixings since they depend on the same couplings as the
Majorana mass terms of the right-handed neutrinos.

\normalsize

\end{titlepage}

\newpage

\setcounter{page}{2} \noindent{\it 1. Introduction. }
 By now the virtues of the supersymmetric (susy) models are well
known, and  in spite of the fact that no supersymmetric particles
have been detected so far, there are hints that supersymmetry may
be a part of the  physical reality, see \cite{glK}. The rich
phenomenology of  supersymmetry  has been widely studied  in
literature, most extensively in the framework of the minimal
supersymmetric standard model (MSSM).

We have recently investigated the phenomenology of a
supersymmetric model based on the left-right symmetric \ssu gauge
theory \cite{HMR}. The left-right symmetric  model (LR-model) is
motivated in particular by the physics of neutrinos:  the solar
\cite{sun} and atmospheric \cite{atmos} neutrino problems as well
as the existence of a hot dark matter component \cite{dark}
suggest that neutrinos should be massive. In the case of  massive
neutrinos, the LR-model is  most natural. It can explain the small
but finite neutrino masses by incorporating the see-saw mechanism
\cite{seesaw}.  However, exactly as is the case with the Standard
Model, the   LR-model is suffering from an unnatural Higgs sector,
which can be made natural by supersymmetrizing the theory. Such a
model was constructed in \cite{HMR}, \cite{bsusylr}.

 In \cite{HMR} the production of a doubly charged higgsino,
$\tilde\Delta^{--}$,
 was studied as a possible experimental signal of the susy
LR-model in different operation modes  ( $e^+e^-$, $e^-e^-$,
$e^-\gamma$ and $\gamma\gamma$) of the next linear collider.  This
particle occurs in the susy LR-model as a member of  the Higgs
triplet superfield. The neutral triplet scalar in this superfield
is responsible for the breaking of the left-right symmetry, and it
also plays a crucial role in the see-saw mechanism due to its
lepton number violating Yukawa couplings with neutrinos. The
higgsino production processes  thus probe the most central
ingredients of the LR-model.

 In the present work we will study the production of a slepton
pair in the susy LR-model with LEP200 and the next linear collider
in mind, {\it i.e.} the process

\be e^+e^-\to\tilde l^+\tilde {l}'^-, \label{reaction}\ee

\noindent  where $\tilde l,\: \tilde l'= \tilde e,\tilde\mu,
\tilde\tau$. The diagrams contributing  are shown in  Fig. 1. In
$s$-channel (Fig. 1a), there is in addition to the MSSM diagrams
also  the diagram  involving the heavy neutral gauge boson $Z_2$.
{}From the Tevatron experiments, the lower limit of the new neutral
gauge boson mass is $m_{Z_2}\ge 310$ GeV \cite{tevatron}, and
therefore we will assume in the following that the  $Z_2$ exchange
contribution can be neglected. Instead of the four neutralinos in
the MSSM model, there are in the susy LR-model all together nine
neutralinos that contribute to the $t$-channel diagram of Fig. 1b.

 The reaction (\ref{reaction}) partly tests the same parts of the
theory as the higgsino production since it is mediated among
others by the doubly charged higgsino (Fig. 1c). It is especially
interesting that one  has a chance to study the flavour changing
couplings of the triplet higgsino, which may exist if neutrinos are
supposed to mix and to oscillate. That is to say, in the susy
LR-model the final state sleptons of (\ref{reaction}) need not be
of the same flavour, in contrast with the situation in the MSSM.

Experimentally, the slepton pair production is possibly one of the
first susy processes to be seen. There are many reasons to that.
Sleptons are supposed to be relatively light among the
superpartners of the standard model particles. Also the decay
pattern of sleptons is simple when compared with many other
supersymmetric particles: in the case of a  light slepton, one
detects a lepton and a large amount of missing energy.

\bigskip

\noindent{\it 2. The supersymmetric left-right model. } The
particle content of  the susy LR-model differs from the particle
content of the MSSM in gauge sector, in Higgs sector, and in having
also a weak isosinglet neutrino superfield which the right-handed
neutrino belongs to. The Higgs sector of the model consists of two
bidoublet superfields transforming under the $SU(2)_L\times
SU(2)_R\times U(1)_{B-L}$ as $({\bf 2},{\bf 2},0)$,

\begin{equation} \begin{array}{c} {\dis\hat\phi
_{u,d}=\matr{cc}\hat\phi_1^0&\hat\phi_1^+\\\hat\phi_2^-&\hat\phi_2
^0 \ematr _{u,d},} \end{array} \end{equation}

\noindent and two right-handed triplet superfields

\begin{equation}  \begin{array}{c}
\dis\hat\Delta=\matr{cc}\frac{1}{\sqrt{2}}\hat\Delta^+&\hat\Delta^{++}\\
\hat\Delta^0&-\frac{1}{\sqrt{2}}\hat\Delta^+\ematr ;\:\:
\dis\hat\delta=\matr{cc}\frac{1}{\sqrt{2}}\hat\delta^-&\hat\delta^{0}\\
\hat\delta^{--}&-\frac{1}{\sqrt{2}}\hat\delta^-\ematr , \end{array}
\end{equation}

\vspace{.1in} \noindent which transform as $({\bf 1},{\bf 3},2)$
and $({\bf 1},{\bf 3},-2)$, respectively. The hatted fields denote
the chiral superfields of the  corresponding field in the ordinary
LR-model. Two bidoublets are needed to have a Kobayashi-Maskawa
matrix not equal an identity, and the second triplet has to be
added to avoid chiral anomalies, which would otherwise appear in
the higgsino sector \cite{bsusylr}. In the gauge sector one has an
extra neutral $\widehat Z_R$  and charged $\widehat W_R^\pm$ gauge
superfields corresponding to the $SU(2)_R$  symmetry.

The superpotential of the supersymmetric left-right model is given
by \cite{HMR},\cite{bsusylr}

\bea W & = & h_{u,ij}^Q \widehat Q_{L,i}^{cT} \widehat \phi_u
\widehat Q_{R,j}  + h_{d,ij}^Q \widehat Q_{L,i}^{cT} \widehat
\phi_d  \widehat Q_{R,j} \nonumber \\ &&+h_{u,ij}^L \widehat
L_{L,i}^{cT} \widehat \phi_u  \widehat L_{R,j}  +h_{d,ij}^L
\widehat L_{L,i}^{cT} \widehat \phi_d  \widehat L_{R,j}
+h_{\DE,ij} \widehat L_{R,i}^{T} i\tau_2 \widehat \DE  \widehat
L_{R,j} \nonumber\\ && + \mu_1 {\rm Tr} (\tau_2 \widehat \phi_u^T
\tau_2 \widehat \phi_d )  +\mu_2 {\rm Tr} (\widehat \DE \widehat
\D ) ,\label{pot} \eea

\noindent where $\widehat Q_{L,i (R,i)}$ denote the left (right)
handed quark superfields of generation $i$ and correspondingly for
leptons $\widehat L_{L,i (R,i)}$. Let us note that we have not
included a left-handed triplet scalar superfield in our theory as
it is unnecessary, unless one wants the superpotential to be
manifestly left-right symmetric.

The Yukawa type interaction terms involving sleptons and higgsinos
and derived from (\ref{pot}) are given by

\vfil\eject

\bea \cL_{ \tilde l-{\rm higgsino\:int.}}  &=& 2h_{\DE ,ij} \tilde
l_{R,i}\tilde\DE^{++}l _{R ,j} + \sqrt{2} h_{\DE ,ij}\tilde l
_{R,i}\tilde\DE^{+}\nu_{R,j}\nonumber \\ && -\tilde l _{L,i}^*
(h_{u,ij}\tilde\phi^0_{2u}+h_{d,ij}\tilde\phi^0_{2d})  l_{R,j}
-\bar l _{L,i} (h_{u,ij}\tilde\phi^0_{2u}+h_{d,ij}
\tilde\phi^0_{2d}) \tilde l  _{R,j}\nonumber\\ &&-\tilde l _{R,i}
(h_{u,ij}\tilde\phi^+_{1u}+h_{d,ij}\tilde\phi^+_{1d})
\bar\nu_{L,j} -\tilde l _{L,i}^*
(h_{u,ij}\tilde\phi^-_{2u}+h_{d,ij} \tilde\phi^-_{2d}) \nu_{R
,j}.\label{Yukawa} \eea The interaction terms  involving sleptons
and gauginos are in turn given by

\bea \cL_{\tilde l  -{\rm gaugino\: interaction}}  &=&  i g_L
\left( -\frac 1{\sqrt{2}}l_L^-\lambda_L^0\tilde l_L^* +
\nu_L\lambda_L^-\tilde l^*_L +l_L^-\lambda_L^+\tilde\nu_L^* +
\nu_L\lambda_L^0\tilde\nu_L^*
 \right) \\ \nonumber && + i g_R \left( \frac
1{\sqrt{2}}l_R^+\lambda_R^0\tilde l_R^* + \nu_R\lambda_R^+\tilde
l^*_R +l_R^+\lambda_R^-\tilde\nu_R^* -
\nu_R\lambda_R^0\tilde\nu_R^*
 \right) \\ \nonumber && +\frac {i g_{B-L}}{2} \left( -\nu_L
\lambda^0_{B-L} \tilde\nu^*_L -l_L\lambda^0_{B-L} \tilde l^*_L
+l_R^+\lambda^0_{B-L} \tilde l^*_R+\nu_R\lambda^0_{B-L}
\tilde\nu^*_R \right) +{\rm h.c.} \eea

\noindent where $\lambda^\pm_{L(R)}$ and $\lambda^0_{L(R)}$ are
the $SU(2)_{L(R)}$ gauginos and $\lambda_{B-L}^0$ is the
$U(1)_{B-L}$ gaugino. In the gaugino part of the Lagrangian there
are no flavour changing interactions while in (\ref{Yukawa}) the
Yukawa coupling constants, the $h$'s, need not be diagonal.

The neutral gauginos and higgsinos mix with the mixing matrix $N$,

\be \tilde \chi^0_{\alpha}  = \sum_{\stackrel {\scriptstyle {\beta
={\rm gauginos},}}{\scriptstyle{{\rm higgsinos}}}} N_{\alpha\beta}
\psi^0_{\beta}. \ee

\noindent The $\tilde\chi_{\alpha}^0$'s form nine physical
Majorana particles. Similarly the charged gauginos and higgsinos
mix with the  two mixing matrices $C^\pm_{\alpha\beta}$,

\be \tilde \chi^\pm_{\alpha}  = \sum_{\stackrel {\scriptstyle
{\beta ={\rm gauginos},}}{\scriptstyle{\rm higgsinos}}}
C^\pm_{\alpha\beta} \psi^\pm_{\beta} \ee

\noindent to form five physical Dirac particles. In the four
component notation, the mixing can be written in the  Lagrangian
as follows

\bea \cL_{\tilde l  l  \chi^0} &=& \half g_L f_{l\alpha}^L\bar l
(1+\G_5) \tilde \chi^0_{\alpha} (\cos \theta \tilde l_1 -
\sin\theta \tilde l_2)\nonumber\\ && -\half g_R f_{l\alpha}^R\bar
l  (1-\G_5) \bar\chi^0_{\alpha} (\sin \theta \tilde l_1
+\cos\theta \tilde l_2)+{\rm h.c.}, \label {llchi}\eea

\noindent where the $\tilde l_{1,2}$ are the mass eigenstates of
sleptons with the assumption that $\theta $ is the mixing angle of
the left- and right-sleptons. In the  following we will assume
that $\theta\ll 1$ and identify $\tilde l_1$ with the left-slepton
$\tilde l_L$ and  $\tilde l_2$ with the right-slepton $\tilde
l_R$. Furthermore, we will assume that the two states are
degenerate in mass, $m_{\tilde l_L}=m_{\tilde l_R}$. They differ
in the opposite chiral structure of their interactions, and they
can  thus be distinguished in experiment for example by using
polarized beams. The mixing factors $f_{l\alpha}^{L,R}$ appearing
in (\ref{llchi}) are given by

\bea f_{l\alpha}^L &=& N_{\alpha 1}+{\frac {g_{B-L}}{g_L}}
N_{\alpha 3}\: ,\nonumber\\ f_{l\alpha}^R &=& N_{\alpha
2}^*+{\frac {g_{B-L}}{g_R}} N_{{\alpha 3}}^*.\label{vmix} \eea

\noindent We have here   neglected the contribution from the
doublet higgsinos as their couplings  are proportional to the
respective Yukawa coupling constants and are therefore small.

The experimental signals of the process (\ref{reaction}) depend on
the decay products of the slepton. These have been studied in the
case of the supersymmetric left-right model  in \cite{HMR}. Light
sleptons decay to leptons and neutralinos: $\tilde l_{L,R}\to
l\tilde\chi^0_i$. If the left-sleptons $\tilde l_L$ are heavier
than one or more of the charginos, then also  the decays $\tilde
l^-_L\to \nu\tilde\chi^-$ might have a large rate.  The
right-sleptons $\tilde l_R$  will also decay, if they are heavy
enough,  to the doubly charged higgsino: $\tilde l^-_{R}\to
l^+\tilde\Delta^{--}$. The other decay modes are kinematically
suppressed. If the lightest supersymmetric particle is one of the
neutralinos, the final state will consist of charged leptons and
some missing energy.

Let us now go on to discuss how does the reaction (\ref{reaction})
probe the various  parameters of the superpotential (\ref{pot}).
The Yukawa couplings $h_{u,ij}^L$ and $h_{d,ij}^L$ of the
bidoublet Higgs fields are known to be small. Therefore this
contribution is negligible, and consequently the dependence of the
cross section on the bidoublet mass parameter  $\mu_1$  is almost
nonexistent. There is no amplitude involving the coupling
$\tilde\Delta^0 \tilde l e $, since that would violate the lepton
number. Thus there is no dependence on the $\mu_2$-parameter
either in neutralino changing graph. Only the gaugino part of the
neutralinos contribute to the selectron pair production.

On the other hand, amplitudes with the $\tilde\Delta^{++} \tilde l
e $ vertex do exist. This is important, since this makes it
feasible to study  by the slepton pair production the
intergenerational coupling of the triplet Higgs to  leptons. The
only diagram, where the higgsino dependence is large, is the graph
including the $\tilde\Delta^{\pm\pm }$ (Fig. 1c). Since this is a
u-channel process,  it is possible to separate its contribution to
the reaction $e^+e^-\to \tilde e^+\tilde e^-$  from that of the
t-channel neutralino diagrams by investigating angular
distributions of the final state particles. The angular
distribution will differ from that in the MSSM offering  a
definitive signal of the left-right symmetric susy theory. This
will  show up, for example,  in the polarization asymmetry of the
final state particles.

Even more definite signals of the susy LR-model would be the final
states which break the separate lepton number conservation, {\it
i.e.} $\tilde e\tilde\mu$, $\tilde e\tilde\tau$ and $\tilde
\mu\tilde\tau$. Such final states are forbidden in the MSSM. In
the susy LR-model they occur in reactions mediated by the triplet
higgsino $\tilde\Delta^{++}$, and their strength is set by the
coupling constants $h_{\Delta,ij}$  ($i\neq j$).

The constants $h_{\Delta,ij}$ appear also in the Majorana mass
terms of the right-handed neutrinos,
$h_{\Delta,ij}\langle\Delta^0\rangle\nu_{i,R}\nu_{j,R}\equiv
M_{ij}\nu_{i,R}\nu_{j,R}$, and are therefore reflected in neutrino
mixing and oscillation. The neutrino mixing is not, however,
barely determined by the triplet Higgs coupling. Indeed, the
masses of light left components of neutrinos are given by the
matrix

\be m=-m_DM^{-1}m_D^T, \ee

\noindent where also the Dirac mass matrix $m_D$ is in general
non-diagonal, and hence the dependence on  $h_{\Delta,ij}$ is
generally highly non-trivial. Nevertheless, if one assumes
quark-lepton symmetry or assumes that there exists a Grand Unified
Theory where the LR-model is embedded, one is able to relate $m_D$
with the quark mass matrix. In that case the information available
from the non-diagonal slepton production would allow one to
estimate neutrino mixing and  thereby to test the various mixing
schemes relevant {\it e.g.} for the solar and atmospheric neutrino
problems.

On the other hand, the mixings and masses of the predominantly
right-handed heavy neutrinos are in the first approximation given
barely by the Majorana couplings $h_{\Delta,ij}$.

For the MSSM, the formulas for the  cross section of the slepton
pair production have been given  elsewhere, \eg in \cite{BFM}.
These formulas can be adapted to the supersymmetric LR-model with
obvious modifications taking  into account the proper gauge
couplings and the mixing of the neutralinos as given in eq.
(\ref{vmix}). \bigskip

\noindent{\it 3. Results and discussion.}
 The selectron pair production in supersymmetric LR-model has a
large cross section when compared with the corresponding process
in the MSSM. This is due to two factors, firstly the number of
gauginos is larger and secondly the triplet higgsino contribution
is large, though dependent on the unknown triplet higgsino
coupling to the electron and selectron.

In Fig. 2 we present the total cross section of the selectron pair
production as a function of the selectron mass $m_{\tilde e}$
($=m_{\tilde e_L}=m_{\tilde e_R}$) for a fixed triplet higgsino
mass $m_{\tilde\Delta}$ (the cross sections depend  rather weakly
on the value of $m_{\tilde\Delta}$). Here and in what follows we
have taken $h_{\Delta,ij}=0.3$. Fig. 2a corresponds to the
situation at LEP200 with $\sqrt{s}=200$ GeV (here
$m_{\tilde\Delta}= 110$ GeV) and Fig. 2b at a linear collider
with  $\sqrt{s}=1$ TeV ($m_{\tilde\Delta}= 300$ GeV). We have
assumed that the left-selectron and the right-selectron are not
identified, the plotted cross section corresponding to the sum
$\sigma(e^+e^-\to \tilde e^+_L\tilde e^-_L)+\sigma(e^+e^-\to
\tilde e^+_R\tilde e^-_R)+ 2\sigma(e^+e^-\to \tilde e^+_L\tilde
e^-_R)$. In the figures we have included two different
supersymmetric LR-models, namely one with the soft gaugino masses
$m_{\lambda_{i}} =1$ TeV (LRM I) and another one with
$m_{\lambda_{i}} =200$ GeV (LRM II). In  both cases we have taken
$\mu_1=\mu_2= 200$ GeV. In the model LRM I the lightest
neutralinos consist mainly of higgsinos, in LRM II mainly of
gauginos. The cross sections in the two cases differ slightly, and
the reason for that can be  understood: in LRM I the $t$-channel
processes are supressed since the higgsino couplings are small, in
LRM II there is no such suppression.

For comparison we have plotted in Fig. 2 also the corresponding
cross section in  the minimal supersymmetric standard model. The
curve MSSM I(II) corresponds to the choice $m_{\lambda_i}=1$ TeV
($200 $ GeV), $\mu=200$ GeV. As one can see, the  cross sections
in the susy LR-model are systematically appreciably larger than
in the minimal supersymmetric standard model.

As mentioned, the most intriguing difference between the susy
LR-model and the minimal susy standard model with respect to the
slepton pair production is the existence of  the $u$-channel
process of Fig. 1c. This reaction, mediated by the SU(2)$_R$
triplet higgsino, occurs only for a right-handed electron and a
left-handed positron, whereas in the $s$- and $t$-channel
processes all chirality combinations may enter. Use of polarized
beams  could therefore give us more information of the triplet
higgsino contribution. Assuming that the decay mode $\tilde e\to
e\tilde\chi^0_1$ is dominant, in Fig. 3 we present the angular
distribution of the final state electron $e^-$ for $\sqrt{s}= 1$
TeV, $m_{\tilde e}=M_{\tilde\Delta}= $ 300 GeV in our two models
LRM I and LRM II. In model LRM I the $t$-channel contributions are
suppressed since the light neutralinos are mainly consisting of
the  higgsinos. In Fig. 3a we present the angular distribution in
the case the electron is right-handedly and the positron is
left-handedly polarized ($P_{+-}$) and in the opposite case
($P_{-+}$). The  distribution $P_{+-}$ is larger and it is
slightly peaked in the backward direction because of the
$u$-channel contribution. In the model LRM II (Fig. 3b) there is
for the both polarization combinations a forward peak. In the case
where the electron has right-handed polarization the forward peak
is, however, less prominent, because there is a large backward
enhancement due to  the $u$-channel reaction.

The dependence of the angular distributions on the mass of the
triplet higgsino $\tilde\Delta^{--}$ shows up clearly in the
polarization asymmetry of the final state electrons in the cascade
process $e^+e^-\to \tilde e^+\tilde e^- \to
e^+e^-\tilde\chi^0_1\tilde\chi^0_1$. As shown in Fig. 4, the
longitudinal polarization asymmetry of the cross section, \be
A_{||}=\frac{\sigma (+,-)-\sigma (-,+)}{\sigma (+,-)+\sigma (-,+)},
\ee (the first and second sign in parenthesis indicate the
longitudinal polarization of the initial state electron and
positron, respectively) is for those final state electrons
originating in right-selectron decays quite sensitive on
$M_{\tilde\Delta}$ (see the curves denoted by R). Particularly
strong this effect is in the model LRM II (Fig. 4b) due to the
interference of the $t$- and $u$-channel contributions, while in
the model LRM I (Fig. 4a), where the $t$-channel is suppressed,
the  $M_{\tilde\Delta}$ dependence is less striking. In Fig. 4 we
have also presented the asymmetry for electrons from the
left-selectron decays  (curve L), which does not depend on the
triplet higgsino since $\tilde e_L$ does not couple with $\tilde
\Delta$. Of course, if the origin of the final state electrons is
not determined the  curves L and R should be added up. We have
assumed in Fig. 4 that the dominant decay channel of selectrons is
$\tilde e\to e\tilde\chi^0_1$, where the neutralino
$\tilde\chi^0_1$ is the lightest supersymmetric particle.

In the MSSM with the unification assumption the right-selectron is
lighter than the left-selectron \cite{heavyeL}. If only $\tilde
e_R$'s are produced, the difference between the MSSM and the
supersymmetric left-right model would be especially large, more
than an order of the magnitude, since in the MSSM  there are  no
SU(2)$_R$ gauginos and in the supersymmetric left-right model the
right handed higgsino gives an extra contribution to the right
slepton pair production.

Finally,  the  cross section of the  pair production of smuons and
staus are in general expected to be smaller than that of
selectron pair production, since the neutralinos do not contribute.
On the other hand the cross sections are in general larger than in
the case of the MSSM because of the nondiagonal couplings of the
triplet higgsinos.

\bigskip \noindent{\bf Ackowledgement.} This work has been
supported by the Academy of Finland.

\newpage {\bf FIGURE CAPTION}

\noindent {\bf Figure 1.} Feynman diagrams for the slepton pair
production in the supersymmetric left-right model.

\noindent {\bf Figure 2.} The total cross section
$\sigma(e^+e^-\to \tilde e^+_L\tilde e^-_L)+\sigma(e^+e^-\to
\tilde e^+_R\tilde e^-_R)+ 2\sigma(e^+e^-\to \tilde e^+_L\tilde
e^-_R)$ as a function of the selectron mass $m_{\tilde l}$ (a) for
the collision energy $\sqrt s=200$ GeV and triplet higgsino mass
$M_{\tilde\Delta}= 110$ GeV, (b) for $\sqrt s=1$ TeV,
$M_{\tilde\Delta}= 300$ GeV. LRM I (II) refer to  two
supersymmetric left-right models and MSSM I (II) to two versions
of the minimal supersymmetric Standard Model described in the text.

\noindent {\bf Figure 3.} The angular distribution of the final
state electron in the cascade process $e^+e^-\to \tilde e^+\tilde
e^-\to e^+e^-\tilde\chi^0_1\tilde\chi^0_1$ for $\sqrt s=1$  TeV,
$M_{\tilde\Delta}= m_{\tilde e}=300$ GeV in the model (a) LRM I,
(b) LRM II. $P_{+-}$ corresponds to the case where the incoming
electron has positive and the incoming positron has negative
longitudinal polarization, and $P_{-+}$ corresponds to the
opposite case.

\noindent {\bf Figure 4.} The longitudinal polarization asymmetry
of the final state electrons from right-selectron decays (curves
R) and from left-selectron decays (curve L) as a function of the
collision energy a) in the model LRM I, b) in the model LRM II. It
is assumed that $\tilde e\to e\tilde\chi^0_1$  is the dominant
decay mode, where the neutralino $\tilde\chi^0_1$ is the lightest
supersymmetric particle. The three curves denoted I, II and III
correspond to the triplet higgsino masses $M_{\tilde\Delta}= 300$
GeV$, 500$ GeV and $800$ GeV, respectively. It is assumed
$m_{\tilde e_L}=m_{\tilde e_R}= 60 $ GeV.


\begin{thebibliography}{99}

\bibitem{glK} G. L. Kane, {\it Is the world supersymmetric?  Do we
already know?}, preprint UM-TH-93-10 (1993), and references
therein.



\bibitem{HMR} K. Huitu, J. Maalampi, M. Raidal, HU-SEFT R 1993-16
(1993).




\bibitem{sun}K. Lande et al., in Proc. XXVth Int. Conf. on High
Energy Physics, eds. K.K. Phua and Y. Yamaguchi (World Scientific,
Singapore, 1991);\\ K.S. Hirata et al., Phys. Rev. Lett. 66 (1990)
1301;\\ K. Nakamura, Nucl. Phys. (Proc. Suppl.) B31 (1993);\\ A.I.
Abazov et al., Phys. Rev. Lett. 67 (1991) 3332;\\ V.N. Gavrin, in
Proc. XXVIth Int. Conf. in High Energy Physics (Dallas 1992), to
appear.

\bibitem{atmos} K.S. Hirata et al., Phys. Lett. B280 (1992) 146;\\
D. Casper et al., Phys. Rev. Lett. 66 (1993) 2561.


\bibitem{dark} M. Davis, F.J. Summers and D. Schlegel, Nature 359
(1992) 393; A.N. Taylor and M. Rowan-Robinson, ibid. 396.


\bibitem{seesaw} M. Gell-Mann, P. Ramond and R. Slansky, in {it
Supergravity}, eds. P. van Niewenhuizen and D. Z. Freedman (North
Holland 1979);\\ T. Yanagida, in Proceedings of {it Workshop on
Unified Theory and Baryon Number in the Universe}, eds. O. Sawada
and A. Sugamoto (KEK 1979).


\bibitem{bsusylr}   M. Cvetic and J. Pati, Phys. Lett. B (1984)
57;\\ Y. Ahn, \PL{149 B} (1984) 337;\\ R. M. Francis, M. Frank, C.
S. Kalman, \PR{D 43} (1991) 2369.

\bibitem{tevatron} F. Abe {\it et al.}, CDF Collaboration, Phys.
Rev. Lett. 68 (1992) 1463.


\bibitem{BFM} A. Bartl, H. Fraas, W. Majerotto, Z. Phys. {\bf C
34} (1987) 411.

\bibitem{heavyeL} L.E. Ibanez and C. Lopez, Phys. Lett. 126 B
(1983) 54, and Nucl. Phys. B 233 (1984) 511.

 \end{thebibliography}
\end{document}